\documentclass[onecolumn]{osa-article}

\journal{osajournal}

\articletype{Research Article}
\usepackage{amsmath}
\usepackage{amsfonts}
\usepackage{graphicx}
\usepackage{bm}
\usepackage{physics}
\usepackage{lipsum}
\usepackage{blindtext}
\usepackage{cuted}
\usepackage{flushend}
\usepackage{lineno}
\usepackage[utf8]{inputenc} 

\begin{document}
\title{Enhancing Near-Field Optical Tweezers by Spin-to-Orbital Angular Momentum Conversion}

\author{Edgar Alonso Guzmán,\authormark{1} Alejandro V. Arzola\authormark{1,*}}
\address{\authormark{1}Instituto de F\'isica, Universidad Nacional Aut\'onoma de M\'exico, C.P. 04510, Cd. de M\'exico, M\'exico}

\email{\authormark{*}alejandro@fisica.unam.mx} 

\begin{abstract} 
Near-field patterns of light provide a way to optically trap, deliver and sort single nanoscopic particles in a wide variety of applications in nanophotonics, microbiology and nanotechnology. Using rigorous electromagnetic theory, we investigate the forces and trapping performance of near-field optical tweezers carrying spin and orbital angular momenta. The trapping field is assumed to be generated by a total internal reflection  microscope objective at a glass-water interface in conditions where most of the transmitted light is evanescent. We find novel aspects of these tweezers, including the possibility to rotate and stably trap nanoscopic beads. More importantly, we show that, under near-field conditions, the contributions of of spin and orbital angular momenta to the rotation of small particles are almost equivalent, opening the possibility to cancel each other when they have opposite sign. We show that these conditions result in optimal optical trapping, giving rise to extremely effective optical tweezers for nanomanipulation, having both circular symmetry and relatively weak rotation. 
\end{abstract}
%

\section{Introduction}

Optical tweezers are broadly used in several fields of physics and biology to deliver objects at will and to explore and understand mechanics at micro and nanoscale \cite{gieseler2020optical, favre2019optical, gao_optical_2017, ashkin_observation_1986, xu2021optomechanical}. These techniques rely on the predominant restoring forces arising from the strong intensity gradients of a laser beam that is tightly-focused close to the diffraction limit, typically by a high numerical-aperture lens (NA $\gtrsim 1.2$) \cite{ashkin_observation_1986}. Standard configurations use the propagating components of the field, limiting the minimum characteristic length of the light-intensity pattern to the order of the wavelength and hence limiting the maximum achievable gradient force for a fixed power. Moreover, these gradient forces acting on subwavelength particles drastically decay as the size of the particle decreases, as the radius to the cube, making their stable confinement against thermal diffusion more difficult or unpractical. For instance, in standard optical tweezers a fraction of a milliwatt of laser power may trap a micronsize particle, but a relatively high power of several tens of milliwats may be required to fully trap a particle with a size below $100\,$nm, accelerating thermal damaging by light absorption  \cite{ashkin_observation_1986}. 

An alternative to optically manipulate subwavelength and nanoscopic particles is the use of near-field components of the electromagnetic field that overcome the diffraction limit \cite{gao_optical_2017}. These techniques have pushed the limits of optical trapping to very small scales, opening the possibility to manipulate, sort or deflect and deliver nanoscopic particles, such as metallic nanoparticles, viruses and macromolecules \cite{bouloumis2020far, hu2020near,erickson2011nanomanipulation,bradac_nanoscale_2018}. In this respect, plasmonic waves in metallic nanostructures, such as nanocavities, metallic waveguides or photonic crystals have been used to produce extremely confined electric fields for nanoparticle manipulation \cite{bouloumis2020far,hu2020near, yoo2018low,erickson2011nanomanipulation, bradac_nanoscale_2018, zaman2019near, berthelot2014three}. Enhanced electric fields with dielectric nanoantennas have also been implemented to manipulate particles \cite{xu2018optical}. In a different approach, subwavelength confinement of light has been accomplished by means of evanescent fields generated by Total Internal Reflection (TIR) at a dielectric interface, for instance by dielectric wave guides, prisms or high numerical-aperture objectives that enable incident angles at the interface larger than the critical angle \cite{liu2021all,emile2020nanoscale,xiang2020optical,li2018living,mohammadnezhad_evanescent_2017,yoon_optical_2010, marchington_optical_2008,brambilla2007optical,gu2004laser,nieto2004near, kawata1992movement, vsiler2008surface}. The latter produces a near-field optical tweezers in the neighborhood of a flat interface typically using a TIR microscope objective \cite{yoon_optical_2010,gu2004laser}.  

In contrast to the patterns generated by propagating fields, near-field patterns exhibit a subwavelength structure, particularly along the direction $z$ perpendicular to the interface or structure. While the intensity of focused propagating fields decay as the inverse of the square of the distance along the propagating axis ($z^{-2}$), evanescent fields exhibit an exponential decay dependence. Although not as strong as plasmonic traps, near-field trapping with all-dielectric configurations has the advantage of preventing thermal heating of the surrounding environment which could be of paramount importance in some applications to preserve biological samples \cite{erickson2011nanomanipulation,liu2021all, xiang2020optical, yoo2018low}. Moreover, near-field trapping by TIR can be easily combined with standard microscope techniques such as fluorescent microscopy, and it is more versatile, in the sense that it does not require fabricated nanostructure \cite{liu2021all, emile2020nanoscale, yoon_optical_2010}.  

Another dynamical property of light that plays an utterly important role in interaction with matter is the angular momentum. In optical tweezers this give rise to non conservative forces making probe samples to spin or rotate about the beam axis  \cite{kotlyar2020spin, garcia2018high, arzola2019spin, zhao2007spin, volke2002orbital, simpson1997mechanical, he1995direct}. The orbital angular momentum per photon (OAM) is $\ell\hbar$, where $\ell$ is an integer known as the topological charge, and the spin angular momentum per photon is $\sigma\hbar$, with $\sigma=\pm1$ for left-hand and right-hand circular polarization. When light is tightly focused to a non-paraxial regime, a fraction of the energy of the beam with  circular polarization and topological charge $\ell$ is converted into an axial vortex with topological charge $\ell\pm1$, depending on the handedness of the circular polarization \cite{bliokh2015spin}. This subtle phenomenon may have important consequences in the behavior of beads trapped in optical tweezers. The radius of rotation of the beads as well as the angular speed may be reduced or increased depending on the relative direction of the circular polarization with respect to the topological charge \cite{arzola2019spin, zhao2007spin}. In contrast to propagating fields where the spin-orbit interaction is a subtle effect, recently it has been shown that this phenomenon has a significant impact in evanescent fields \cite{bliokh2014extraordinary}, particularly in connection with optical trapping \cite{mei2016evanescent}.

Therefore, the existing methods for generating near-field patterns contribute  to improve and extend the capabilities of standard optical tweezers to manipulate nanoscopic particles. However, some important aspects of near-field patterns produced by total internal reflection remain unexplored, such as evanescent fields that have both spin and orbital angular momenta and, specifically, the effect of strong spin-orbit interaction on the stability of optical trapping.

\begin{figure}
\centering
\includegraphics[width=3in]{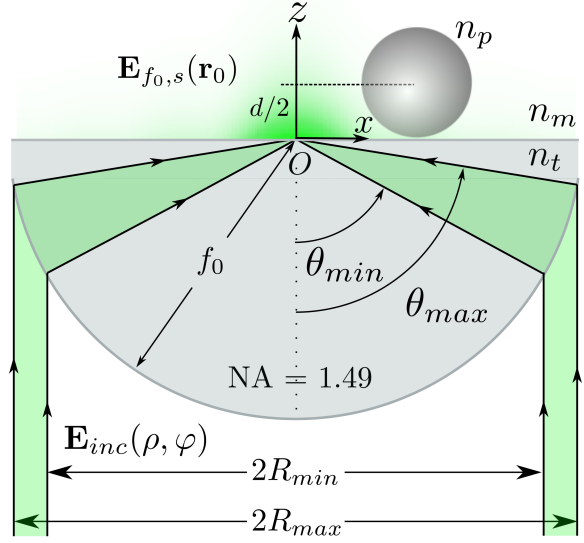}
\caption{Geometrical configuration of the near-field optical tweezers with spin and orbital angular momenta. An annular laser beam with amplitude ${\bf E}_{inc}$, azimuthal phase $\ell\phi$ and polarization $\sigma$, passes through a TIR microscope objective with numerical aperture NA=1.49 (gray-color hemisphere) and is focused at the interface in $O$. The annular shape is defined by the radii $R_{min}$ and $R_{max}$.  The objective, with focal length $f_0$, refracts the rays of the input beam radially to the center of the hemisphere. The bundle of refracted rays form a conic shell defined by the angles $\theta_{min}$ and  $\theta_{max}$. The interface is defined by the media with refractive indices $n_t=1.51$ and $n_m=1.33$. An evanescent field with amplitude ${\bf E}_{f_0,s}$ is generated above the interface by TIR at the medium with refractive index $n_m$. This highly z-confined field enables to trap nanometric particles with refractive index $n_p$ and diameter $d$.}\label{fig:fig1}
\end{figure}
  
Here we present a novel and thorough numerical study of the effect of spin and orbital angular momentum in near-field optical tweezers based on a TIR microscope objective. We use the Richards and Wolf representation of the focused field at the interface and the generalized Lorenz-Mie theory to compute the optical forces \cite{jones2015optical, richards1959electromagnetic}. With this in hand, we analyze the force field emerging from the evanescent fields, putting especial attention to the cases where the trap renders an stable equilibrium point. To initially explore the behavior and  stability of these traps, we analyze their orbital rotation, owing angular momentum transfer, and stiffness as a function of the particle size. Additionally, we explore the trapping efficiency following two criteria: a new quantity called efficacy and the potential depth, allowing us to have a fair comparison on their performance with respect to the input laser power. We show that the properties and performance of these evanescent traps can be drastically modified depending on the combination of orbital and spin angular momenta showing that the maximum trapping performance is given when the incoming beam has a topological charge with $\ell=\pm1$ pointing oppositely to the spin in such a way that $\ell\sigma=-1$. As Ashkin et al. pointed out in their seminal work for a single-beam optical trap, given a wavelength there is a threshold of laser power below which small particles remain undamaged for a certain time \cite{ashkin_observation_1986}. In this respect, the results presented here may imply a significant reduction in input laser power and, hence, thermal heating, to safely trap nanometric particles.	

%

\section*{Near-Field Optical Tweezers}

In our simulation, we use the Richards and Wolf theory of a focused laser beam \cite{richards1959electromagnetic, jones2015optical, novotny2012principles} to model a laser beam with an annular profile that is tightly focused by a lens with numerical aperture NA=1.49 (TIR Objective) at a glass-water interface, as it is schematically illustrated in Fig~\ref{fig:fig1}. The components of the angular spectrum of the input field are redirected by the lens to the center of a sphere with radius equal to the focal distance $f_0$. The focal point $O$ is located on the interface between the media with refractive indices $n_t$ and $n_m$, with $n_t>n_m$. Therefore, to generate a predominantly evanescent field at the interface, we assume that the internal and external radii of the ring of light have to take the values $R_{\textrm{min}}=f_0\sin\theta_{min}$ and $R_{\textrm{max}}=f_0\sin\theta_{max}=f_0\textrm{NA}/n_t$, with $\theta_{min}$ equal to the critical angle $\theta_c=\sin^{-1}(n_m/n_t)=61.74^{\circ}$ for total internal reflection. Using NA=1.49, $f_0=3.33\,$mm (NA and focal length of a commercial Nikon microscope objective), $n_t=1.51$ and $n_m=1.33$, we obtain $R_{min}=2.94\,$mm and $R_{max}=3.29\,$mm. 

The spatial complex amplitude of the electric field entering into the objective is

\begin{equation}\label{eq:Ein}
    \mathbf{E}_{inc}(\rho,\varphi)=\frac{E_0}{\sqrt{2}}e^{-{\rho^2/w^2}}e^{i\ell\varphi}(\hat{\mathbf{x}}+i\sigma\,\hat{\mathbf{y}}),
\end{equation}
for $R_{min}\leq \rho\leq R_{max}$, and zero otherwise. The beam is expressed in cylindrical coordinates ($\rho$, $\varphi$). $E_0$ is the electric field amplitude at the center of a gaussian profile with radius $w=R_{max}=3.29\,$mm. The wavelength of the laser beam in vacuum is $\lambda_0=532\,$nm. This beam has both orbital and spin angular momenta, given by the azimuthal phase $\ell\varphi$ with topological charge $\ell$, and by the circular polarization, respectively. The effective power passing through the optical system is

\begin{equation}\label{Eq:power}
P_{eff}=\frac{\pi}{4} c\varepsilon_0|\mathbf{E}_{0}|^2w^2\left[e^{-2R^2_{min}/w^2}-e^{-2R^2_{max}/w^2}\right],
\end{equation}
where $c$ and $\varepsilon_0$ are the speed of light and the electric permittivity of vacuum. The evanescent field passing through the interface at position $\mathbf{r}_0$ is \cite{novotny2012principles}

\begin{align} 
&\mathbf{E}_{f_0,s}(\mathbf{r}_0) = \frac{ik_tf_0e^{ik_tf_0}}{2 \pi}  \int \limits_0^{2\pi} \int \limits_{\theta_{min}}^{\theta_{max}}\textbf{E}_{\text{ff,s}}(\theta,\varphi) e^{i\textbf{k}_s \cdot \mathbf{r}_0} \sin \theta \dd \theta\dd\varphi,\label{eq:Ef0}\\
&\text{where} \nonumber \\
&\mathbf{E}_{\text{ff,s}}(\theta,\varphi) = \sqrt{\frac{\cos\theta}{n_t}}\left( t_{\parallel}(\theta)[\mathbf{E}_{inc}(\rho,\varphi) \cdot \bm{\hat{\rho}}]\,\bm{\hat{s}} + t_{\bot}(\theta)[\mathbf{E}_{inc}(\rho,\varphi) \cdot \bm{\hat{\varphi}}]\,\bm{\hat{\varphi}} \right),\label{eq:Eff}\\
&\mathbf{k}_s=(k_{s,x},k_{s,y},k_{s,z})=k_t\sin\theta\cos\varphi\,\hat{\mathbf{x}}+k_t\sin\theta\sin\varphi\,\hat{\mathbf{y}}+k_t\sqrt{n_s^2/n_t^2-\sin^2{\theta}}\,\hat{\mathbf{z}},\nonumber \\
&\textbf{k}_t =(k_{t,x},k_{t,y},k_{t,z}) = k_t\sin\theta\cos\varphi\,\hat{\mathbf{x}}+k_t\sin\theta\sin\varphi\,\hat{\mathbf{y}}+k_t\cos\theta \,\hat{\mathbf{z}}, \nonumber \\
&\hat{\mathbf{s}}=\cos{\theta}_s\sin{\varphi}\,\hat{\mathbf{x}}+\cos{\theta}_s\cos{\varphi}\,\hat{\mathbf{y}}-\sin{\theta_s}\,\hat{\mathbf{z}},\nonumber\\ 
&\bm{\hat{\rho}}=\cos{\varphi}\,\hat{\mathbf{x}}+\sin{\varphi}\,\hat{\mathbf{y}},\nonumber\\
& \bm{\hat{\varphi}} = -\sin \varphi \,\hat{\mathbf{x}} + \cos \varphi \,\hat{\mathbf{y}} \nonumber \\
&\text{and} \ k_t=kn_t=\frac{2\pi}{\lambda_0}n_t.\nonumber 
\end{align}

\noindent The incident and transmitted angles, $\theta$ and $\theta_s$, satisfy the Snell's law at the interface, $n_t\sin \theta = n_m\sin \theta_s$. The  intensity of the transmitted light is $I(\mathbf{r}_0)=c\varepsilon_m|\mathbf{E}_{f_0,s}(\mathbf{r}_0)|^2/2n_m$. The force exerted by this field on a dielectric sphere with diameter $d$ and refractive index $n_p$ is computed by means of the generalized Lorenz-Mie theory (see Methods for more details) \cite{jones2015optical}.

\begin{figure*}
\centering
    \includegraphics[width=5in]{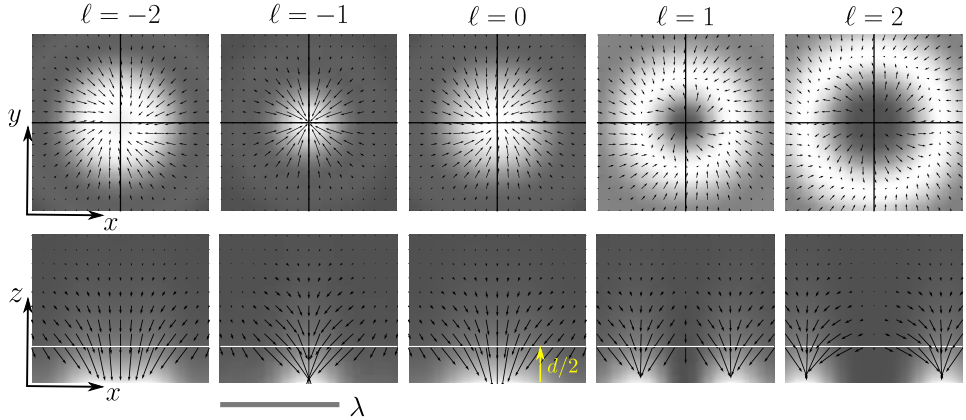}
    \caption{Intensity profiles and force fields in near-field optical tweezers with different topological charges ($\ell$) and left-hand circular polarization ($\sigma=+1$).  The wavelength in the surrounding medium is $\lambda=400$nm ($n_m=1.33$). Top row: Transverse (XY) intensity profiles at the interface. Bottom row: Corresponding axial (XZ) intensity profiles. The black arrows indicate the force acting on a particle of size $d=240\,$nm and refractive index $n_p=1.59$. Using the effective power $P_{eff}=2\,$mW, the maximum intensities in each configuration are (from left to right): $12.50\text{mW}/\mu\text{m}^2$, $27.54\text{mW}/\mu\text{m}^2$, $14.85\text{mW}/\mu\text{m}^2$, $9.70\text{mW}/\mu\text{m}^2$ and $7.99\text{mW}/\mu\text{m}^2$. The corresponding maximum lengths of the arrows for each case are, along XY plane: $0.12 $pN, $0.24 $pN, $0.13 $pN, $0.09 $pN, and $0.10 $pN, and along XZ plane: $0.24 $pN, $0.42$pN, $0.33 $pN, $0.19 $pN, and $0.15$pN.}
    \label{fig:fig2}
\end{figure*}

In practice, the annular profile of the beam at the entrance of the microscope objective could be achieved by means of an annular filter, as those for TIRF microscopy, located at the back aperture of the objective. Another more efficient approach would be based on refraction by axicon lenses in a configuration designed to generate a parallel annular beam \cite{lei2013long, yoon_optical_2010,volke2002orbital}. An annular beam generated by means of a focused Bessel beam is also feasible \cite{khonina2020bessel}. The azimuthal phase with topological charge $\ell$ can be addressed by means of an spatial light modulator or by a spiral phase plate, and the state of polarization can be defined by a quarter wave plate \cite{arzola2019spin}. 

\section*{Evanescent fields and trapping}
Figure~\ref{fig:fig2} shows in gray scale the intensity of the evanescent patterns at the interface for topological charges $\ell=-2,-1, 0, 1$ and $2$ at the XY plane (top) and along the axial plane XZ (bottom). In this article we are only considering left-hand circular polarization ($\sigma=+1$ or $+\hbar$ per photon). The intensity is drastically affected by the combination of orbital angular momentum and spin at the entrance of the objective. The spin to orbit conversion occurs when the circularly polarized field at the entrance pupil is refracted and projected along the z axis by the high numerical-aperture lens and is the responsible for the asymmetric behavior with respect to $\ell$ illustrated in Fig.~\ref{fig:fig2}. It was previously pointed out that the axial field under these conditions resembles a vortex with topological charge $\ell+1$ for $\sigma=+1$ or $\ell-1$ for $\sigma=-1$ \cite{bliokh2015spin, arzola2019spin,  zhao2007spin}. For positive topological charges, in the same direction as the polarization, the vortex singularity becomes evident by the dark spot at the center and a bright doughnut shape typically appearing in focused beams carrying OAM \cite{bliokh2015spin}. On the other hand, for the negative values  of $\ell$, opposite to the spin, the vortex singularity disappears. 
Contrary to the patterns generated by propagating fields, these patterns exhibit a subwavelength structure. The intensity pattern generated with the combination $\ell=-1$ and $\sigma=+1$ has the narrowest spot size, approximately half the wavelength, anticipating an aspect that will become important to improve optical trapping at the nanoscale. Moreover, as expected in evanescent fields, the XZ profiles (bottom row in Fig.\ref{fig:fig2}) exhibit an abrupt decay of the intensity along z, ideally with an exponential dependence.

The black arrows in Figure~\ref{fig:fig2} indicate the force field exerted on a probe bead with $d=240\,$nm and refractive index $n_p=1.59$ (polystyrene).  In all cases we can see the strong restoring forces along $z$, enabling confinement at the interface where the intensity of the evanescent field the highest. In the transverse patterns (XY plane), the contributions of conservative and rotational (non conservative forces), are readily perceived by the way the arrows converge (diverge), and rotate around the central equilibrium point. The deviation of the arrows from the radial direction show the rotational nature of the force field (see arrows near the solid black lines along $x$ and $y$ axes), which is a consequence of the angular momentum of the field \cite{garcia2018high, arzola2019spin,ruffner2012optical, zhao2007spin}. It is important to stress that a particle in a stable equilibrium point does not rotate, but if it is driven out from equilibrium, for example, by external perturbations or thermal fluctuations, it will gain rotational motion around the beam axis. The force fields with $\ell=1, 2$ exhibit an unstable equilibrium point at the center and a limit cycle with rotation along the most intense circumference, following the anticlockwise rotation of the angular momentum. On the other hand, the force fields corresponding to $\ell=-2,-1$ and $0$ show an stable equilibrium point at the center of the pattern. The force field for $\ell=-2$ shows a weak clockwise rotation and the force field for $\ell=0$ shows a very similar rotation pattern but in the opposite direction, both of them following the rotation direction of the total angular momentum $\ell+\sigma=-1,1$. Noteworthy, the rotational component of the force field is almost null or negligible in the case of $\ell=-1$, in accord with $\ell+\sigma=0$. These facts suggest that the non conservative forces arising from the orbital angular momentum with $\ell=\pm1$ and from the spin angular momentum $\sigma=\pm1$, defined at the entrance pupil of the microscope objective, are almost equivalent at the interface. We deduce that this unique phenomenon is owing to the high efficiency of spin to OAM conversion under evanescent conditions.
At this point, we can infer that the optical trap with the combination of parameters $\ell=-1$ and $\sigma=+1$ (or equivalently with $\ell=1$ and $\sigma=-1$), which exhibits the narrowest spot size and imperceptible rotation, may represent an extraordinary improvement of optical traps to manipulate nanometric particles. As it was shown previously, larger topological charges either positive or negative give rise to limit cycles enabling subwavelength optical spanners \cite{mei2016evanescent}. 

\section*{Stiffness and stability} 
\begin{figure}
\centering
     \includegraphics[width=3in]{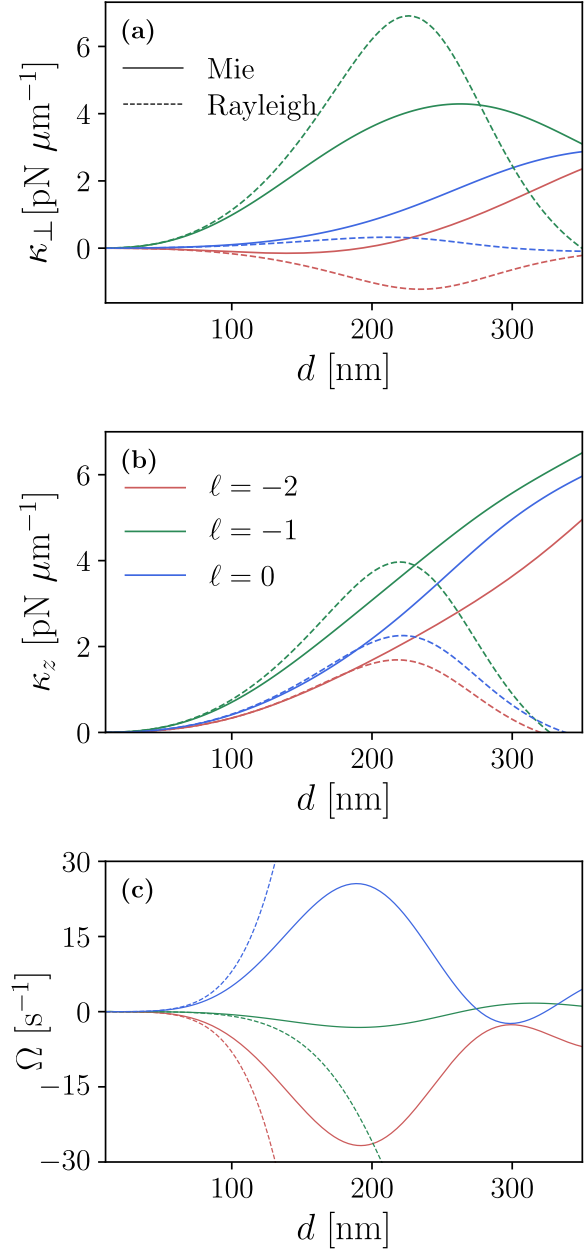}
     \caption{Stiffness and rotation frequency of the force field in the neighborhood of the equilibrium point in the evanescent optical tweezers as a function of the size of the particle, $d$, for the  topological charges $\ell=-2, -1$ and $0$. The input beam has left-hand circular polarization ($\sigma=+1$). The center of the particle is located at a distance $d/2$ away from the interface. Solid lines represent the results from the  generalized Lorenz-Mie theory while dashed lines are the results obtained with the Rayleigh approximation.  (a) Stiffness along x, (b) stiffness along z and (c) rotation frequency with respect to the beam axis in the neighborhood of the equilibrium point. The refractive index of the particle is $n_p=1.59$ and the effective power passing through the optical system is $P_{eff}=2\,$mW.}
    \label{fig:fig3}
 \end{figure}

As it was previously described, the strong gradient forces along $z$ pull the particle towards the interface, setting its position at $z=d/2$ due to the counteracting force of the hard wall. At this plane, the particle is attracted to the center $(x,y)=(0,0)$ or repelled from it depending on its size and the topological charge. Here, we will drive our attention to the cases with $\ell=-2, -1$ and $0$ (see Fig.~\ref{fig:fig2}) where the equilibrium point is predominantly stable for a large range of particle sizes, meaning that the particle is stiffly trapped at this point instead of having a limit cycle. The linear form of the force field in the neighborhood of the equilibrium point is $\mathbf{F}_{eq}(x,y)=\mathbf{J}_0(0,0)(x,y)^{T}$, where $\mathbf{J}_0$ is the 2D Jacobian, which can be considered as the result from the superposition of an harmonic conservative potential and a non-conservative rotational force field \cite{volpe2007brownian,garcia2018high}. Given the circular symmetry of the evanescent patterns shown in Fig.~\ref{fig:fig2}, the 2D Jacobian can be directly written as 
\begin{equation}\label{eq:Jacobian}
\mathbf{J}_0=\begin{bmatrix}
           -\kappa_\perp&-2\pi\gamma\Omega\\
           2\pi\gamma \Omega&-\kappa_\perp
             \end{bmatrix},
\end{equation}
where the parameter $\kappa_{\perp}=-\partial_x F_x$ is the transversal stiffness of the trap (due to the symmetry $\kappa_{\perp}=-\partial_y F_y$) and $\Omega=-(2\pi\gamma)^{-1}\partial_y F_x$ is a frequency, providing an information on the amount of rotation of the particle about the beam axis in the neighborhood of the equilibrium point. Likewise, the stability along $z$ can be characterized by $\kappa_{z}=-\partial_z F_z$, which does not precisely correspond to that of an harmonic potential, since the particle is actually equilibrated by the counteracting force of the hard wall at the interface. For our purposes, we define the stiffness of the trap along this direction considering only $z \gtrsim d/2$. The friction coefficient is taken as $\gamma=3\pi \nu d$ with viscosity $\nu=1\,\text{m}\,\text{Pa}\,\text{s}$, which corresponds to the ideal condition of an spherical particle in water where the hydrodynamic and electrostatic interactions with the glass-water interface are ignored. A more detailed description should take into account hydrodynamic corrections (Faxen's law) as well as the DLVO interaction with the interface \cite{jones2015optical}. The components of the Jacobian are plotted in Fig.~\ref{fig:fig3} as a function of the particle diameter for a fixed effective power $P_{eff}=2\,$mW. For the sake of comparison, we also include the results computed by means of the Rayleigh (dipole) approximation \cite{albaladejo2009scattering, novotny2012principles, jones2015optical}. 

These results confirm the relatively small rotation of particles in evanescent fields with topological charge $\ell=-1$ and $\sigma=+1$, and confirm that this configuration represents the most efficient optical trap for tiny beads, with sizes $d\lesssim 350\,$nm, considering the transverse and axial axis. The field with $\ell=-2$ exhibits the lowest efficiency, since the transverse stiffness turns very weak and negative for beads below $200\,$nm, meaning that the center in those cases represents an unstable equilibrium point. We are not including the cases with $\ell=1$ and $2$ shown in Fig.~\ref{fig:fig2}, or larger, since we are interested in the analysis of stable equilibrium points for small particles. It may be the case that these topological charges, or even larger ($|\ell|>2$), may exhibit stable equilibrium points for some particles with larger sizes than those explored here, but this is out of the scope of this article. It is clear that the Rayleigh theory, applicable only for very small particles, renders good predictions for particles smaller than $100\,$nm. Contrary to the typical monotonic incremental behavior of the force as the radius of the particle to the cube for small particles, in this case the curves decay for large particles because the particle is set $d/2$ away from the interface and hence it receives less laser energy as its size increases.  

\begin{figure}
\centering
     \includegraphics[width=3in]{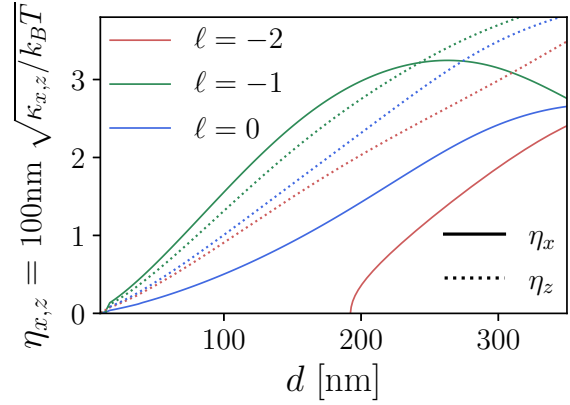}
     \caption{Trapping efficacy of the near-field optical traps as a function of the diameter of the particle, $d$.  Solid and dotted lines indicate the efficacy along $x$ and $z$, respectively. This parameter is defined as the ratio of a characteristic trapping length to the standard deviation of the Brownian particle in the harmonic well with stiffness $\kappa_{x,z}$. The trapping length was chosen as $s_h=\lambda/4=100\,$nm. The particle has a refractive index $n_p=1.59$ and we set $P_{eff}=2\,$mW for the power.}
   \label{fig:fig4}
 \end{figure}

In addition to the stiffness of the optical trap, the stability of a particle in the potential well depends on thermal diffusion, which is responsible for taking the particle out of the equilibrium and eventually push it out of the trap. To evaluate the effectiveness of near-field optical tweezers, we can consider the strength of confinement against the effect of thermal diffusion only in the region where the trap is harmonic. The standard deviation of the position of the particle under this condition is defined by the equipartition theorem as $\sigma_x=\sqrt{k_B T/\kappa_x}$. Assuming a harmonic  potential in a neighborhood with characteristic scale $s_{h}\approx\lambda/4=100\,$nm, we define the efficacy of the trap as 
\begin{equation}\label{eq:efficacy}
\eta_{x,z}=\frac{s_{h}}{\sigma_{x,z}}=s_{h}\sqrt{\frac{\kappa_{x,z}}{k_B T}},    
\end{equation}
which allows us to have a clearer idea of the stability of the particle with diameter $d$ in the optical trap at fixed temperature, $T$, and input power, $P_{eff}$. In other words, this parameter tells us how strong is the particle trapped at a temperature $T$ in a small region $s_{h}$ when we use the total power $P_{eff}$ in the trap.  We can think of $\eta_{x} > 1$ as a reasonable condition for effectively trapping the particle. Figure~\ref{fig:fig4} shows $\eta_{x,z}$ as a function of the diameter for an input power $P_{eff}= 2\,$mW. Since $\eta_{x}$ is proportional to the square root of the power via $\kappa_x$, an increase of $\eta_{x}$ to an arbitrary value $q$, requires a power increase by $q^2$, this is $P_{eff}|_{q}=q^2\,(2\,\text{mW}/\eta^2_{x,z})$, using the computed value of $\eta_{x,z}$ with $P_{eff}=2\,$mW as a reference. For instance for $\ell=-1$ (see Fig.~\ref{fig:fig4}), a particle with diameter $d=94\,$nm has $\eta_x=1.48$, while a particle with $d=240\,$nm has $\eta_x=3.2$. In order to trap the former particle with the same efficacy as the latter, we would require $P_{eff}|_{\eta_x=3.2}=9.34\,$mW of power entering into the system instead of the $2\,$mW originally used to compute Fig.~\ref{fig:fig4}. It is also clear from Fig.~\ref{fig:fig4} that $\eta_{z}$ is comparable to $\eta_x$ for $\ell=-1$.

\begin{figure}
    \centering
    \includegraphics[width=3in]{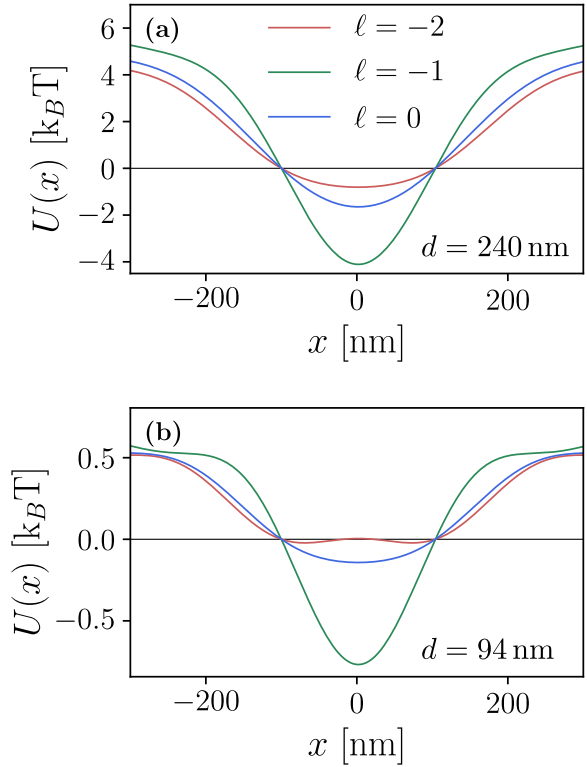}
    \caption{Comparison of the potentials in the near-field tweezers for two particle sizes for $\ell=-2, -1$ and $0$ and $\sigma=+1$. (a) potential for a particle with $d=240\,$nm and (b) for a particle with $d=94\,$nm. The refractive index of the particle is $n_p=1.59$ and the effective power passing through the optical system is $P_{eff}=2\,$mW.}
    \label{fig:fig5}
\end{figure}

Figure~\ref{fig:fig5} shows the potentials along $x$  for a large, $d=240\,$nm, and for a small particle, $d=94\,$nm, for a fixed power $P_{eff}=2\,$mW. In order to have a better comparison of these potentials,  all of them are plotted with their zero value fixed at $x=s_h=100\,$nm. This helps to visualize the stable neighborhood and does not affect the analysis of the potential. As it was previously pointed out, the most effective near-field traps correspond to the combination of parameters $\ell=-1$ and $\sigma=+1$,  which is confirmed again in these figures with the deepest potentials. The particle with $d=240\,$nm is tightly trapped with about $4.1\,k_BT$ in the neighborhood of the equilibrium point, whereas the particle with $d=94\,$nm is poorly trapped with only $0.77\,k_B T$. In the latter case, the particle will explore non-harmonic regions of the near-field trap and easily abandon the equilibrium neighborhood due to thermal fluctuations . Using this criterion, increase of the power up to $P_{eff}=4.1\,(2\,\text{mW}/0.77)=10.6\,$mW will improve the trap with a potential depth in the equilibrium neighborhood of $4.1\,k_BT$, equal to the one of the $240\,$nm. Since we are only considering the harmonic neighborhood of the potential, this criterion results equivalent to the one defined by $\eta$ once we assume that the particle is well trapped with $4.1\,k_BT$ in the neighborhood of the equilibrium point. We believe that the efficacy $\eta_{x,z}$, which scales as the square root of the power, gives a clearer picture to compare the stability of optical traps and could eventually help to compare optical traps generated with very different structured-patterns of light. 

\section*{Conclusions}
Based on Richards and Wolf theory and on the generalized Lorenz-Mie theory to estimate optical forces, we performed a study of the optical forces exerted by near-field optical traps with spin and orbital angular momenta on subwavelength dielectric particles. We assumed that the near-field was generated at a glass-water interface by total internal reflection of a field focused with a TIR microscope objective. We considered an objective lens with a fixed numerical aperture (NA=1.49) enabling an annular opening at the entrance with radii $R_{max}$ and $R_{min}$. This annular opening allowed a high level of total internal reflection. In future studies it would be interesting to explore how these parameters modify the intensity gradients and optical forces in near-field tweezers, in connection with Ref.~\cite{mei2016evanescent}. At the entrance plane of the objective, the total spin and orbital angular momenta are set by the polarization and the topological charge, respectively.  We first described the intensity profile of the evanescent field as a function of the topological charge when we had left-hand circular polarization ($\sigma=+1$). Afterwards, we computed the optical forces, the stiffness at the equilibrium points and the potentials for interesting configurations. We found it useful to define the efficacy of the trap as the ratio of the characteristic scale in the intensity profile to the standard deviation of the Brownian motion of the particle in the harmonic potential, enabling us to define the stability of the Brownian particle in the optical tweezers. 

Overall, we showed that the spin and orbital angular momenta in the input field play a very important role in the stability of the probe beads. There are cases where the vortex structure dominates the profile and the particles can be pushed towards a limit cycle depending on their size. The cases with the lowest total angular momentum, {\it i.e.} with $\ell+\sigma=-1,0,1$ show a bright spot in the center with stable equilibrium points for most particle sizes. We also showed that the profiles with zero total angular momentum, {\it i.e.} with $\ell=-1$ and $\sigma=+1$, or equivalently with $\ell=1$ and $\sigma=-1$, exhibit the narrower intensity profile with almost no rotational component. The force exerted by this configuration turns out to be very effective to tightly trap nanometric particles, putting forward a way to increase the efficacy several times with respect to other near-field optical traps. Our results contribute to the understanding and design of novel mechanisms with structured evanescent light, which may significantly improve all-dielectric near-field optical tweezers for nanoparticle manipulation. The future experimental realization of this efficient, rotationless and circularly symmetric near-field tweezers would allow to strongly trap nanoscopic particles along $x$ and $z$ axes with few milliwatts of laser power, providing a robust tool in the fields of nanomanipulation and spin-orbit interaction. Moreover, future experimental realizations of these tweezers could open an avenue to address novel properties of optical forces in confined fields, such as non-conservative gradient forces and  other reactive effects that manifest with particles with both electric and magnetic response \cite{xu2019azimuthal, xu2020kerker, nieto2021reactive}.

\section*{Methods}
{\bf Computation of optical forces}. The force exerted by the electromagnetic field on a spherical particle of diameter $d$ at position $\mathbf{r}=(x,y,z)$ is 
\begin{equation} 
\textbf{F}(\textbf{r}) =  \oint_\Gamma \left< \textbf{T} \right> \cdot \bm{\hat{r}}_s \ \dd \Gamma
\label{eq:ForcesMieMaxwellTensorInt}
\end{equation}
\noindent where $\left<\mathbf{T}\right>$ is the time-averaged Maxwell stress tensor evaluated on an spherical surface $\Gamma$  surrounding the particle and $\bm{\hat{r}}_s$ is the unitary vector normal to the surface $\Gamma$. This tensor is given in therms of the electric field, $\mathbf{E}$, and the magnetic induction, $\mathbf{B}$, as

\begin{equation}
\langle\mathbf{T}\rangle= \frac{1}{2} \varepsilon_m \text{Re}\left\{ \textbf{E} \otimes \textbf{E}^{*} + \frac{c^2}{n_m^2} \textbf{B} \otimes \textbf{B}^{*} - \frac{1}{2}\left( |\textbf{E}|^2 + \frac{c^2}{n_m^2}|\textbf{B}|^2 \right)\textbf{I}\right\}.
\end{equation}

\noindent The parameters $c$, $\varepsilon_m$ and $n_m$ are the speed of light in vacuum, the permittivity and the refractive index of the surrounding medium, respectively.  

The electromagnetic field at $\mathbf{r}_s$ on the surface $\Gamma$ follows from the sum of the incident and scattered fields: $\textbf{E}(\textbf{r}_s)=\textbf{E}_i(\textbf{r}_s)+\textbf{E}_s(\textbf{r}_s)$ and $\textbf{B}(\textbf{r}_s)=\textbf{B}_i(\textbf{r}_s)+\textbf{B}_s(\textbf{r}_s)$. The scattered fields ($\textbf{E}_\mathrm{s}(\textbf{r}_s)$ and $\textbf{B}_\mathrm{s}(\textbf{r}_s)$) are given by the multipole expansion in therms of the vector spherical harmonics ($\textbf{H}_{lm}^{(1)}$ and $\textbf{H}_{lm}^{(2)}$) \cite{jones2015optical,novotny2012principles} 

\begin{equation}
\begin{split}
\textbf{E}_s(r_s,\bm{\hat{\mathbf{r}}}_s) & = \sum_{l=0}^{\infty} \sum_{m=-l}^{l} \sum_{p=1}^2 \mathcal{A}_{s,lm}^{(p)}\textbf{H}_{lm}^{(p)}(kr_s,\bm{\hat{\textbf{r}}}_s), \\
i\frac{c}{n_m}\textbf{B}_{s}(r_s,\bm{\hat{\textbf{r}}}_s) & = \sum_{l=0}^{\infty} \sum_{m=-l}^{l} \sum\limits_{\substack{ p = 1 \\ p\neq p'}}^{2} \mathcal{A}_{s,lm}^{(p)}\textbf{H}_{lm}^{(p')}(kr_s,\bm{\hat{\textbf{r}}}_s)
\end{split}
\label{eq:ScatteringFieldsExpansion}
\end{equation}

\noindent where $k$ is the wave number in the medium with refractive index $n_m$. The coefficients $\mathcal{A}_{s,lm}^{(1)} = -b_l\mathcal{W}_{s,lm}^{(1)}$ and $\mathcal{A}_{i,lm}^{(2)} = -a_l\mathcal{W}_{i,lm}^{(2)}$ are related with Mie coefficients $a_l$ and $b_l$ \cite{jones2015optical}
. Hence, the coefficients $\mathcal{W}_{\mathrm{i},lm}^{(p)}$ for a particle with the center at $\mathbf{r}$ are given by \cite{jones2015optical}

\begin{equation}\label{eq:WCoefficientsFocusingInterface}
\mathcal{W}_{i,lm}^{(p)}(\mathbf{r}) = \frac{ik_tf_0e^{ik_tf_0}}{2 \pi}   \int \limits_0^{2\pi}\int \limits_{\theta_{min}}^{\theta_{max}} \textbf{E}_{\text{ff,s}}(\theta,\varphi) W_{\mathrm{i},lm}^{(p)}e^{i(k_{t,z}-k_{s,z})z_i}e^{i\textbf{k}_s \cdot \mathbf{r}} \sin \theta  \dd \theta \dd\varphi,
\end{equation}

\noindent where $\textbf{E}_{\text{ff,s}}(\theta,\varphi)$ is defined by Eq.~\eqref{eq:Eff}. $\textbf{k}_t $ and $\textbf{k}_s$ are the wave vectors in each side of the interface, and $z_i$ is the axial position of the focal point with respect to the interface as if there were no interface. Here we chose $z_i=0$. The angles  $\theta_{min}$ and $\theta_{max}$ are defined by the internal and external rays of the incident annular beam at the focal point (see Fig.~\ref{fig:fig1}). The coefficients $
W_{i,lm}^{(p)}(\bm{\hat{\textbf{e}}},\bm{\hat{\textbf{k}}}) = 4\pi i^{p+l-1}(-1)^{m+1} \bm{\hat{\textbf{e}}} \cdot \textbf{Z}_{l,-m}^{(p)}(\bm{\hat{\textbf{k}}}) $  
are the multipole coefficients of the expansion of a plane wave in therms of the vector spherical harmonics ($\textbf{J}_{lm}^{(1)}$ and $\textbf{J}_{lm}^{(2)}$) with polarization $\bm{\hat{\mathrm{e}}}$ \cite{borghese2007scattering,jones2015optical}. $\textbf{E}_{inc}$ is the electric field impinging the lens (see Fig.~\ref{fig:fig1} and Eq.~\eqref{eq:Ein}), while
$t_{\parallel}$ and $t_{\bot}$ are the coefficients of Fresnel for transmission for p and s polarization. The unitary vectors $\bm{\hat{\rho}}$, $\bm{\hat{\varphi}}$ and  $\bm{\hat{s}}$ are defined in Eq.~\eqref{eq:Eff}.


\section*{Competing interests}
  The authors declare that they have no competing interests.

\section*{Author's contributions}
    A.V.A. developed the concept and supervised the study. E.A.G. implemented the models and performed the simulations. A.V.A. and E.A.G. discussed the results and A.V.A. wrote the draft of the article. All authors revised the final version of the article.
\section*{Acknowledgements}
We thank Mariana Ben\'itez and Karen Volke for critical reading of the manuscript. 

\section*{Funding}
This work was supported by UNAM-PAPIIT IN111919.

\section*{Data availability}
Data underlying the results presented in this paper are not publicly available at this time but may be obtained from the authors upon reasonable request.


\end{document}